# Room Temperature Magnetic Order in Air-Stable Ultra-Thin Iron Oxide

Jiangtan Yuan[1#], Andrew Balk[2#], Hua Guo[1], Sahil Patel[1], Xuanhan Zhao[3], Qiyi Fang[1], Douglas Natelson[3], Scott Crooker[2], Jun Lou[1*]

[1]Department of Materials Science and NanoEngineering, Rice University, Houston, TX 77005, USA
[2]National High Magnetic Field Laboratory, Los Alamos, New Mexico 87545, USA
[3]Department of Physics and Astronomy, Rice University, Houston, TX 77005, USA
[#] Contributed equally to this work

**Certain two-dimensional (2D) materials exhibit intriguing properties such as valley polarization[1], ferroelectricity[2], superconductivity[3] and charge-density waves[4,5]. Many of these materials can be manually assembled into atomic-scale multilayer devices[6,7] under ambient conditions, owing to their exceptional chemical stability. Efforts have been made to add a magnetic degree of freedom to these 2D materials via defects, but only local magnetism has been achieved[8-10]. Only with the recent discoveries of 2D materials supporting intrinsic ferromagnetism have stacked spintronic devices become realistic[11-15]. Assembling 2D multilayer devices with these ferromagnets under ambient conditions remains challenging due to their sensitivity to environmental degradation, and magnetic order at room temperature is rare in van der Waals materials. Here, we report the growth of air-stable ultra-thin epsilon-phase iron oxide crystals that exhibit magnetic order at room temperature. These crystals require no passivation and can be prepared in large quantity by cost-effective chemical vapor deposition (CVD). We find that the epsilon phase, which is energetically unfavorable and does not form in bulk, can be easily made in 2D down to a seven unit-cell thickness. Magneto-optical Kerr effect (MOKE) magnetometry of individual crystals shows that even at this ultrathin limit the epsilon phase exhibits robust magnetism with coercive fields of hundreds of mT. These measurements highlight the advantages of ultrathin iron oxide as a promising candidate towards air-stable 2D magnetism and integration into 2D spintronic devices.**

Iron oxides are abundant in nature and are present in almost every domain on earth, including the atmosphere, biosphere and lithosphere.[16] They are also among the most studied metal oxides, having been applied exhaustively for technological applications such as data storage and catalysis, and biomedical applications such as drug delivery, medical imaging, and cancer treatment.[16] The most common polymorphs of iron oxide are $\alpha$-$Fe_2O_3$ (hematitie), $\gamma$-$Fe_2O_3$ (maghemite), and $Fe_3O_4$ (magnetite), which exist in both bulk and nanoscale forms. In contrast, $\varepsilon$-$Fe_2O_3$ is a rare phase with little natural existence and has only been found at the nanoscale.[17] It has received far less research interest than the other polymorphs, due partially to its difficulty in preparation, as it cannot be grown in bulk. However, $\varepsilon$-$Fe_2O_3$ has a variety of interesting characteristics, such as ferrimagnetism, multiferroicity[18], and a large coercive field[19], motivating investigation into growth techniques and properties.

To date, ε-Fe$_2$O$_3$ has been prepared as nanoparticles with sol-gel methods[19,20], thermal annealing[21] and as thin films with pulsed laser deposition[22], electrodeposition[23], and atomic layer deposition (ALD)[24]. While these important advances have enabled research into the properties of ε-Fe$_2$O$_3$, the morphology of the materials produced in these ways prevents potential incorporation into 2D spintronic devices. Here, we show synthesis of ultrathin, 2D crystals with ambient pressure CVD, finding that 2D crystals of ε-Fe$_2$O$_3$ readily form on both silicon and mica substrates. Electron microscopy and Raman spectroscopy measurements confirm that the crystals are pure ε-Fe$_2$O$_3$, with no detectable amounts of the more common α-Fe$_2$O$_3$ and γ-Fe$_2$O$_3$, for crystals thinner than approximately 100 nm. Furthermore, MOKE magnetometry of individual crystals show that they are magnetically stable, with coercive fields of hundreds of mT. We observe robust hysteresis even in crystals as thin as 7 nm at room temperature in atmosphere. Moreover, our samples of 2D ε-Fe$_2$O$_3$ can be readily transferred from growth substrates in aqueous solutions at ambient conditions to arbitrary substrates without any visible structural changes. Finally, despite the fact that these CVD-grown 2D ε-Fe$_2$O$_3$ crystals are not van der Waals materials, their atomically sharp surfaces and nanoscale thicknesses will allow them to have the potential to be easily integrated with other 2D materials, thereby eliminating the lattice mismatching constraints for design of functional heterostructures. This CVD growth, manipulation, and magnetic study of comparatively large individual crystals is complementary to recent successes in liquid-phase exfoliation of ensembles of nanoscale ultrathin hematite crystals[25].

ε-Fe$_2$O$_3$ is not a layered material. It has a orthorhombic structure with lattice constants $a$ = 5.072 Å, $b$ = 8.736 Å, $c$ = 9.418 Å and belongs to the space group of $Pna2_1$.[17] There are four independent crystallographically nonequivalent iron sites, denoted as Fe$_A$, Fe$_B$, Fe$_C$ and Fe$_D$, occupying the center of either the octahedron or tetrahedron formed by surrounding oxygen atoms (Fig. 1a). We grow samples with standard CVD techniques (see Methods for growth details). Briefly, iron chloride tetrahydrate (FeCl$_2$·4H$_2$O) powder was placed in an aluminum oxide crucible at the center of furnace, while fluorophlogopite mica was used as a growth substrate for most of the samples discussed here. During the growth process, iron chloride tetrahydrate transitions to the vapor phase of iron chloride and water, then iron chloride reacts with water, oxidizes to ε-Fe$_2$O$_3$, and finally deposits on the mica substrate. The growth system is protected using Argon as a carrier gas. Ultra-thin crystals of iron oxide with lateral size of ≈10 μm readily form using this procedure, and are apparent in optical micrographs (Fig. 1b-d) All crystals exhibit hexagonal, half-hexagonal or triangular shapes with sharp edges, showing clear evidence of high crystallinity.

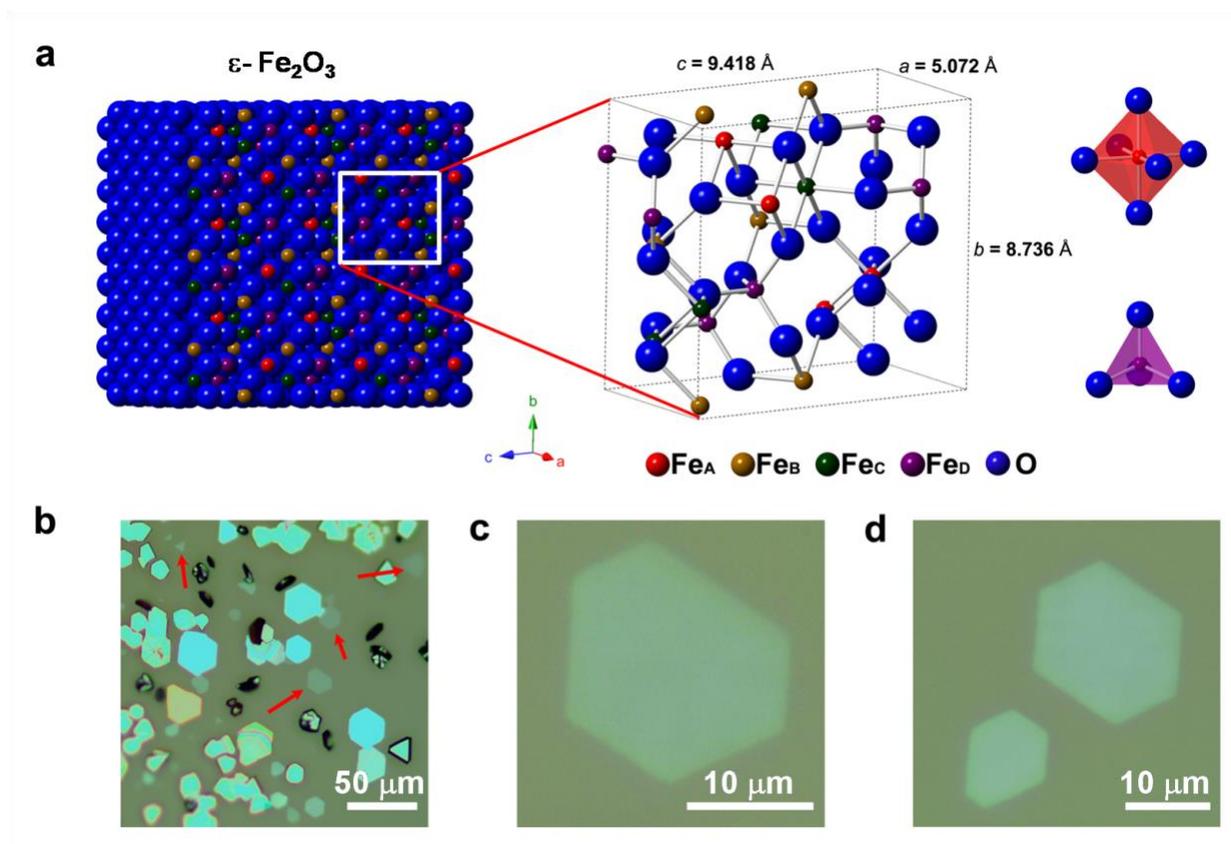

**Figure 1 | Crystal structure and optical images of ε-Fe$_2$O$_3$. a,** Non-layered ε-Fe$_2$O$_3$ has an orthorhombic structure with $a$ = 5.072 Å, $b$ = 8.736 Å and $c$ = 9.418 Å. There are four independent iron sites, denoted as Fe$_A$, Fe$_B$, Fe$_C$ and Fe$_D$. **Inset:** Individual octahedron formed by one center iron atom and six surrounding oxygen atoms, representing the cation coordination of Fe$_A$, Fe$_B$, Fe$_C$, and an individual tetrahedron formed by one center iron atom and four surrounding oxygen atoms, representing the cation coordination of Fe$_D$. **b-d,** Optical images of ε-Fe$_2$O$_3$ crystals grown on mica by CVD. Ultrathin crystals with a lateral size of many μm can be easily found in each batch of growth, as indicated by the red arrows.

Fig. 2a and 2b are atomic force microscopy (AFM) images of two representative crystals with thicknesses of ≈5.8 nm and ≈7.5 nm. The thinnest crystal we have measured is ≈5.1 nm which is only five unit-cell thickness. The 2D Fe$_2$O$_3$ crystals have smooth surfaces and uniform thicknesses, with standard deviation roughness less than 0.5 nm. Micro-Raman study on these samples (Fig. 2g) shows four peaks between 100 cm$^{-1}$ and 200 cm$^{-1}$, which are the characteristic first-order phonon vibration modes M1-M4 of ε-Fe$_2$O$_3$, in agreement with previous literature[20]. In comparison, the most stable bulk phase α-Fe$_2$O$_3$ has no Raman active modes in this range. Spatially resolved Raman mapping (Fig. 2e, f) further suggests uniformity within individual crystals. No detectable second phase was observed.

We also employed transmission electron microscopy (TEM) to gain further structural insight of the as-synthesized 2D Fe$_2$O$_3$. Fig. 2h is a bright-field TEM image of a 2D Fe$_2$O$_3$ polycrystalline flake. The electron diffraction pattern (Fig. 2h inset) from a randomly selected area can be indexed to the orthorhombic symmetry of ε-$Fe_2O_3$ in the [001] zone axis, consistent with the results from Raman spectroscopy. In addition to examining multiple randomly transferred crystals, we performed TEM and Raman analysis on the same ≈ 100 nm thick crystal by breaking it into two halves. The resulted TEM index matches well with that of Raman analysis (characteristic peaks between 100 $cm^{-1}$ and 200 $cm^{-1}$), confirming that Raman spectroscopy is an accurate and rapid way to identify 2D ε-$Fe_2O_3$. In total, we have checked 23 crystals with thicknesses range from 5.1 nm to 260 nm. Of these, 22 crystals show the ε phase (see Extended Data Fig. 1).

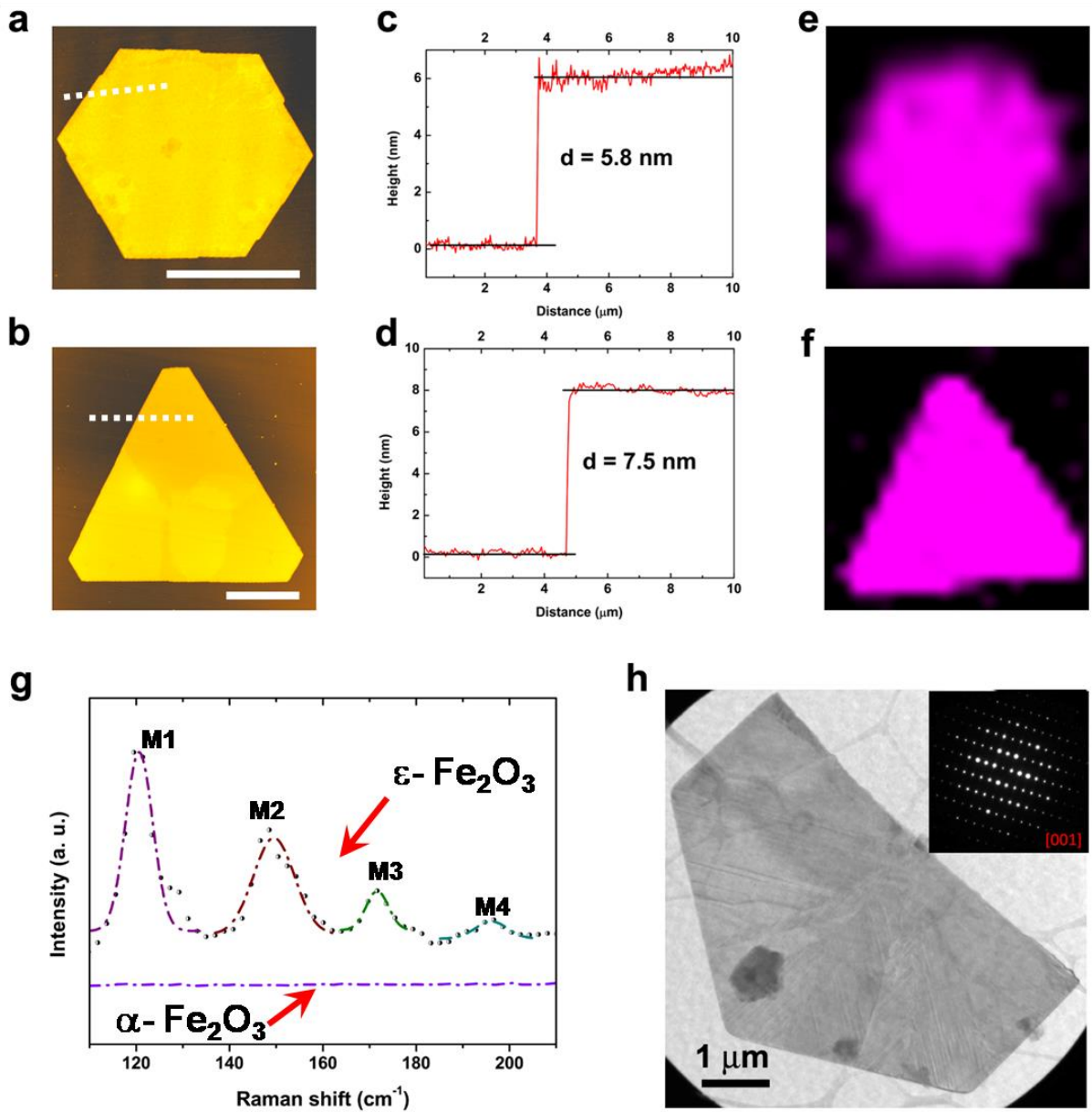

**Figure 2 | Characterization of ε-Fe$_2$O$_3$. a**, **b**, Two representative 2D ε-Fe$_2$O$_3$ crystals with hexagonal and triangular shapes grown on a mica substrate. The well-defined shapes indicate high crystal quality. Scale bars are 5 μm. **c**, **d**, AFM thickness measurements for crystals in **a**, **b**. The thicknesses are 5.8 nm and 7.5 nm, respectively. **e**, **f**, The corresponding spatially resolved Raman mapping for crystals in **a** and **b**. The intensities are the sum of signals between 100 cm$^{-1}$ and 200 cm$^{-1}$, normalized by Raman peak of mica substrate. The homogeneous intensity distributions show that ε-Fe$_2$O$_3$ phase is uniformly distributed within the crystals. **g**, Raman spectra collected from ε-Fe$_2$O$_3$ and α-Fe$_2$O$_3$ crystals on silicon oxide. The four characteristic peaks between 100 cm$^{-1}$ and 200 cm$^{-1}$ represent the first-order phonon modes, namely M1-M4, of ε-Fe$_2$O$_3$, whereas α-Fe$_2$O$_3$ has no active Raman modes in this range. Peaks are fitted by Lorentz functions. **h**, Bright-field TEM image of a 2D ε-Fe$_2$O$_3$ polycrystalline flake transferred from mica substrate. Inset: Electron diffraction pattern which can be indexed to the orthorhombic symmetry of ε-Fe$_2$O$_3$ in the [001] zone axis.

We next probe the magnetic properties of these 2D ε-Fe$_2$O$_3$ crystals with longitudinal MOKE measurement at room temperature (295K). The measurement geometry is shown as inset in Fig. 3a, and further details of the measurement and data processing are described in Methods. Fig. 3a-d show typical hysteresis loops of Kerr rotation $\theta_K$ as a function of magnetic field, $B$, for samples with thickness ranging from 7.5 nm to 50.1 nm. These loops clearly show room-temperature magnetic order in ultra-thin ε-Fe$_2$O$_3$ with well-defined transitions, and coercive fields of hundreds of mT. Noting the difference in the amplitude of the Kerr effect, in particular between the 7.5 nm and 15 nm thick samples (Fig. 3a and Fig. 3c), we investigated 15 other crystals with different thicknesses to check for an influence of thickness on magnetic properties. Of these, all show similar room temperature switching behavior, with a mean coercive field of 290 mT and a standard deviation of 80 mT. These results agree well with previous measurements of nanoparticles[17,21] and thin films of ε-Fe$_2$O$_3$[22]. To check the robustness of the magnetic properties of our samples at room temperature, we plot the coercive field and the amplitude of the magnetic transitions as a function of sample thickness, as measured by AFM. We find that neither the amplitude of the Kerr effect (Fig. 3e) nor the coercive field (Fig. 3f) correlates with sample thickness. This absence of correlation confirms the robustness of magnetic order in these materials.

This observation is in contrast with other magnetic samples. For example, magnetite (Fe$_3$O$_4$) films[26,27] and nanoparticles[28] show a strong dependence of coercive field on sample dimensions. In this regard, our data on 2D ε-Fe$_2$O$_3$ suggests its promising potential in ultra-compact information storage applications.

Electronic transport measurements (see Extended Data Figures 3 and 4) show that the material is highly resistive, with a room temperature resistivity on the order of 100 Ω-m and no observable gate dependence or magnetic field dependence. This is comparable to expectations for the related oxide, hematite[29,30], and implies high sample quality through the lack of doping from vacancies or impurities.

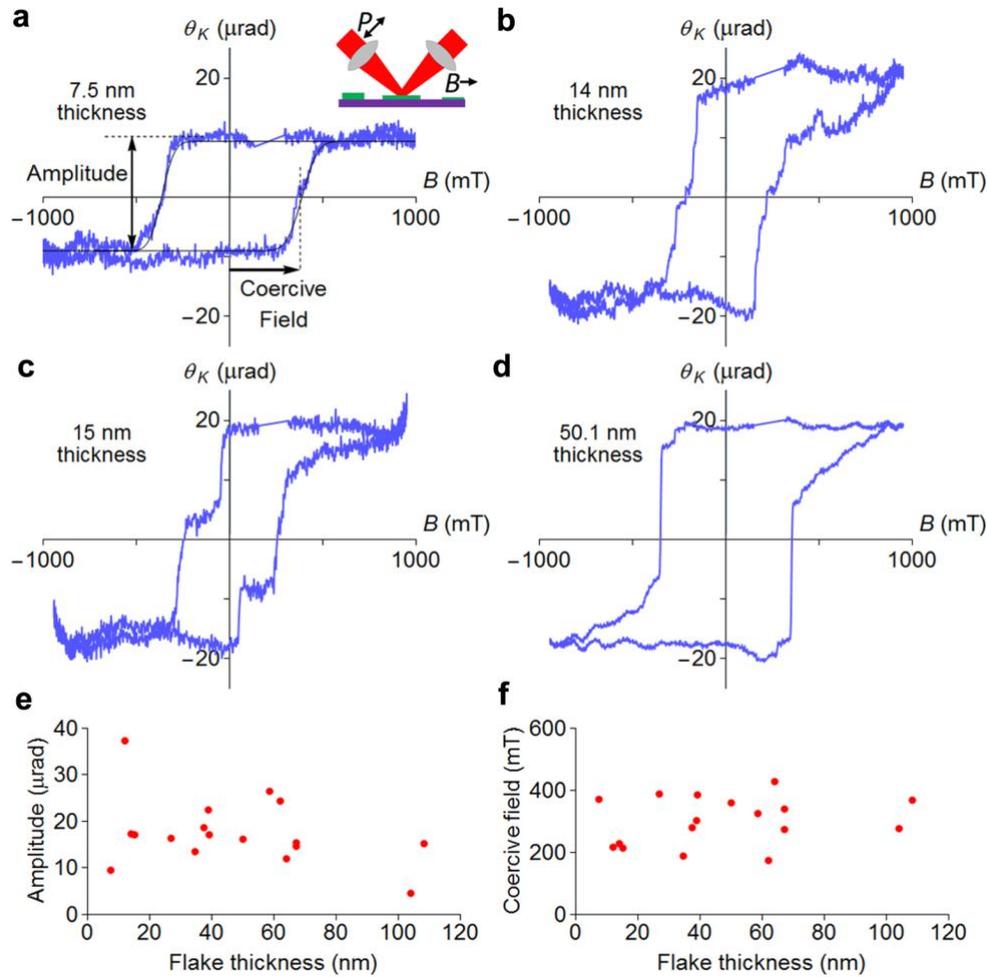

**Figure 3 | Room-temperature magnetic order in nanoscale thickness ε-Fe₂O₃.** **a-d**, Hysteresis loops obtained from crystals with varying thicknesses from 7.5 nm to 50.1 nm, demonstrating magnetic order with symmetric hysteresis and coercive fields ≈ 300 mT. **e**, The amplitude of the Kerr effect plotted against the thickness of the crystals measured by AFM, showing no correlation. **f**, The coercive field plotted as a function of thickness. Coercive fields and amplitudes are measured from the hysteresis loops by fitting the magnetic transitions to error functions. Backgrounds were removed from these hysteresis curves according to the procedure described in the extended data.

One distinctive advantage of 2D ε-Fe₂O₃ is that these crystals have exceptional stability. Fig. 4a is an AFM image of a crystal stored in ambient condition (T = 24.0 °C, RH = 39%) for over three months. No obvious voids or morphology changes were observed. Height profile analyses across the whole sample (Fig. 4b) reveals that the crystal has smooth surface, with a standard deviation roughness of 0.28 nm and sharp edges. Therefore it is not likely that these 2D ε-Fe₂O₃ crystals will be further oxidized, demonstrating extraordinary stability compared to other 2D magnetic materials reported so far.

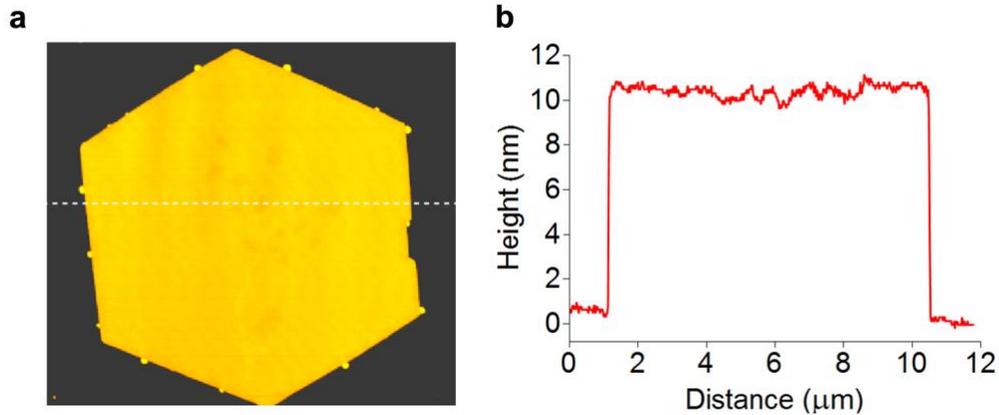

**Figure 4 | Stability of ε-Fe$_2$O$_3$. a**, AFM image of a crystal stored under ambient conditions for over three months. No obvious voids or morphology changes can be observed. **b**, Height profile analysis across the whole crystal indicates that the surface is highly smooth, with a standard deviation roughness of 0.28 nm, and edges are sharp without any signs of degradation.

**Conclusion**

Ultra-thin 2D ε-Fe$_2$O$_3$ can be synthesized by a simple ambient CVD technique. MOKE measurements on individual crystals clearly show that 2D ε-Fe$_2$O$_3$ is magnetically ordered at room temperature with coercive fields of 200 to 400 mT and sharp magnetic transitions, even down to 7.5 nm thickness. Moreover, we find that 2D ε-Fe$_2$O$_3$ will not degrade under ambient conditions for months. Thanks to these useful and unique properties, ε-Fe$_2$O$_3$ will be a promising and distinctive platform to explore magnetism in the 2D limit. It is envisioned that new conceptual devices with novel spin functionalities could be developed through heterostructure engineering with other relevant 2D materials in the near future.

1   Mak, K., McGill, K., Park, J. & McEuen, P. The valley Hall effect in MoS$_2$ transistors. *Science* **344**, 1489-1492, (2014).
2   Chang, K. *et al.* Discovery of robust in-plane ferroelectricity in atomic-thick SnTe. *Science* **353**, 274-278, (2016).
3   Gozar, A. *et al.* High-temperature interface superconductivity between metallic and insulating copper oxides. *Nature* **455**, 782-785, (2008).
4   Xi, X. *et al.* Strongly enhanced charge-density-wave order in monolayer NbSe$_2$. *Nat. Nanotechnol.* **10**, 765-769, (2015).
5   Ritschel, T. *et al.* Orbital textures and charge density waves in transition metal dichalcogenides. *Nat. Phys.* **11**, 328-331, (2015).
6   Novoselov, K., Mishchenko, A., Carvalho, A. & Neto, A. 2D materials and van der Waals heterostructures. *Science* **353**, (2016).
7   Geim, A. & Grigorieva, I. Van der Waals heterostructures. *Nature* **499**, 419-425, (2013).
8   Magda, G. *et al.* Room-temperature magnetic order on zigzag edges of narrow graphene nanoribbons. *Nature* **514**, 608-611, (2014).

9   Gonzalez-Herrero, H. *et al.* Atomic-scale control of graphene magnetism by using hydrogen atoms. *Science* **352**, 437-441, (2016).
10  Seixas, L., Carvalho, A. & Neto, A. Atomically thin dilute magnetism in Co-doped phosphorene. *Phys. Rev. B* **91**, 155138 (2015).
11  Huang, B. *et al.* Layer-dependent ferromagnetism in a van der Waals crystal down to the monolayer limit. *Nature* **546**, 270-273, (2017).
12  Gong, C. *et al.* Discovery of intrinsic ferromagnetism in two-dimensional van der Waals crystals. *Nature* **546**, 265-269, (2017).
13  Bonilla , M. *et al.* Strong room-temperature ferromagnetism in $VSe_2$ monolayers on van der Waals substrates. *Nat. Nanotechnol.*, doi:10.1038/s41565-018-0063-9 (2018).
14  Song, T. *et al.* Giant tunneling magnetoresistance in spin-filter van der Waals heterostructures. *Science*, doi: 10.1126/science.aar4851 (2018).
15  Klein, D. R. *et al.* Probing magnetism in 2D van der Waals crystalline insulators via electron tunneling. *Science*, doi: 10.1126/science.aar3617 (2018).
16  Cornell, R. M. & Schwertmann, U. *The Iron Oxides: Structure, Properties, Reactions, Occurrences and Uses*. Second edn, (WILEY-VCH GmbH & Co. KGaA, 2003).
17  Tucek, J., Zboril, R., Namai, A. & Ohkoshi, S. epsilon-$Fe_2O_3$: an advanced nanomaterial exhibiting giant coercive field, millimeter-wave ferromagnetic resonance, and magnetoelectric coupling. *Chem. Mater.* **22**, 6483-6505, (2010).
18  Gich, M. *et al.* Multiferroic Iron Oxide Thin Films at Room Temperature. *Adv. Mater.* **26**, 4645-4652, (2014).
19  Jin, J., Ohkoshi, S. & Hashimoto, K. Giant coercive field of nanometer-sized iron oxide. *Adv. Mater.* **16**, 1, (2004).
20  Lopez-Sanchez, J. *et al.* Sol-gel synthesis and micro-Raman characterization of epsilon-$Fe_2O_3$ micro- and nanoparticles. *Chem. Mater.* **28**, 511-518, (2016).
21  Taboada, E., Gich, M. & Roig, A. Nanospheres of silica with an epsilon-$Fe_2O_3$ single crystal nucleus. *Acs Nano* **3**, 3377-3382, (2009).
22  Gich, M. *et al.* Epitaxial stabilization of epsilon-$Fe_2O_3$ (00l) thin films on $SrTiO_3$ (111). *Appl. Phys. Lett.* **96**, (2010).
23  Kulkarni, S. & Lokhande, C. Structural, optical, electrical and dielectrical properties of electrosynthesized nanocrystalline iron oxide thin films. *Mater. Chem. and Phys.* **82**, 151-156, (2003).
24  Tanskanen, A., Mustonen, O. & Karppinen, M. Simple ALD process for epsilon-$Fe_2O_3$ thin films. *APL Mater.* **5**, 056104 (2017).
25  Balan, A. P. *et al.* Exfoliation of a non-van der Waals material from iron ore hematite. *Nat. Nanotechnol.*, doi: 10.1038/s41565-018-0134-yy (2018).
26  Moussy, J. *et al.* Thickness dependence of anomalous magnetic behavior in epitaxial $Fe_3O_4$(111) thin films: Effect of density of antiphase boundaries. *Phys. Rev. B* **70**, 174448 (2004).
27  Margulies, D. *et al.* Anomalous moment and anisotropy behavior in $Fe_3O_4$ films. *Phys. Rev. B* **53**, 9175-9187 (1996).
28  Goya, G., Berquo, T., Fonseca, F. & Morales, M. Static and dynamic magnetic properties of spherical magnetite nanoparticles. *Jour. Appl. Phys.* **94**, 3520-3528 (2003).
29  Morin, F. Electrical properties of $\alpha$-$Fe_2O_3$ and $\alpha$-$Fe_2O_3$ containing titanium. *Phys. Rev.* **83**, 1005 (1951).

30  Glasscock, J., Barnes, P., Plumb, I., Bendavid, A. & Martin, P. Structural, optical and electrical properties of undoped polycrystalline hematite thin films produced using filtered arc deposition. *Thin Solid Films* **516**, 1716-1724, (2008).

**Acknowledgements**

The CVD growth, Raman and AFM characterization of $Fe_2O_3$ was supported by the Welch Foundation (C-1716). TEM characterization was supported by a DOE BES grant (DE-SC0018193). MOKE measurements at the NHMFL were supported by the National Science Foundation DMR-1644779. Device fabrication and transport measurements were funded by Rice IDEA support and National Science Foundation DMR-1704264. We thank K. H. and L. Yang for their helpful discussion, C. Sui for assistance with TEM sample preparation, and Panpan Zhou for assistance with device fabrication. ALB acknowledges support of the Los Alamos LDRD program.

**Author Contributions**

J. Y. and J. L. conceived the idea and designed the experiments. J. Y., S. P. and Q. F. conducted the materials growth and Raman, AFM characterization. A. B. and S. C. performed the MOKE measurements. A.B. processed and analyzed the MOKE data. H. G. carried out the TEM characterization and phase index from electron diffraction patterns. X. Z. and D. N. handled the device fabrication for magnetotransport measurements. All authors wrote the manuscript and discussed the results at all stages.

**Author Information**

The authors declare no competing financial interests. Correspondence and requests should be addressed to J. L. (jlou@rice.edu).

**METHODS**

**CVD of 2D $Fe_2O_3$.** Ultra-thin 2D $\varepsilon$-$Fe_2O_3$ was directly grown on both $SiO_2$/Si and fluorophlogopite mica (Changchun Taiyuan Fluorphlogopite Co., China) by CVD using iron chloride tetrahydrate ($FeCl_2 \cdot 4H_2O$) powder (Sigma-Aldrich) as precursor. Prior to growth, mica substrates were freshly cleaved with dimensions about 10 mm x 10 mm. The substrates were placed on top of an aluminum boat which was located in a 2-inch-diameter quartz tube furnace (MTI), with about 0.15g $FeCl_2 \cdot 4H_2O$ powder. The tube furnace was purged with ultra-high purity Ar (99.999%) gas at 50 standard cubic centimeters per minute (sccm) for 10 min to create an inert environment. Then, the furnace was heated to 740 ºC at a rate of 35 ºC/min and held at that temperature for 20 min. Finally, the furnace was opened at 300 ºC and allowed to cool until room temperature was reached. The exhaust was cleaned by a gas collection system containing NaOH solution before it went out to a gas outlet.

**Characterization of 2D Fe$_2$O$_3$.** AFM images of individual flake were collected in ambient conditions using a Bruker Multimode 8 system in tapping mode. Bruker TESPA-V2 n-doped Si cantilevers with a resonance frequency of approximately 320 kHz were used with a scan rate of 0.977 Hz. Line-scan artifacts were removed from the AFM images. Raman spectra were obtained with a Renishaw inVia confocal Raman microscope. The laser spot size was focused with a 50X lens to a diameter of ≈ 1 μm. In sharp contrast to the Raman spectrum of α-Fe$_2$O$_3$ (hematite) where there were no peaks below 200 cm$^{-1}$, ε-Fe$_2$O$_3$ showed four peaks between 100 cm$^{-1}$ and 200 cm$^{-1}$. It was much easier to distinguish these fours peaks when ε-Fe$_2$O$_3$ flakes were deposited on SiO$_2$/Si substrate (Fig. 2g) than of mica due to a large background from mica. The Raman mapping images in Fig. 2 were collected from flakes directly grown on mica. The color intensity is proportional to the ratio between the integrated intensity between 100 cm$^{-1}$ and 180 cm$^{-1}$, and the mica peak intensity at 196 cm$^{-1}$. The existence of ε-Fe$_2$O$_3$ flakes were clearly indicated from the contrast (red or purple from flakes vs. black from mica substrate). TEM bright-field images and electron diffraction patterns were acquired by a JEM-2100F at an acceleration voltage of 200 kV. We prepared the TEM samples using two techniques: For the first, we transferred flakes grown on mica to lacey carbon TEM grids through a PMMA aided liquid transfer process using water instead of NaOH. In this case, two flakes on TEM grids were randomly chosen for phase identification, and both showed ε phase. For the second, we broke one crystal into two halves. One half was moved to a TEM grid for electron diffraction, the other half left for Raman spectroscopy analysis. Both electron diffraction and Raman spectrum results indicated the ε phase, thereby confirming Raman spectroscopy was a reliable and speedy method for phase identification.

**Magneto-optical Kerr effect measurement.** We performed magnetometry with MOKE in the longitudinal geometry. The mica substrate with ε-Fe$_2$O$_3$ flakes was first attached to a sample holder mounted between the poles of an electromagnet. For field excitation, the electromagnet was linearly ramped back and forth between 950 mT and -950 mT at a rate of 440 mT/s. Simultaneously, laser light with a wavelength of 632.8 nm and a nominal total power of 1.5 mW was focused on individual flakes at an angle of incidence of 45°, resulting in a nominal spot diameter of 6 μm on the sample. Focus was confirmed with an optical microscope aimed at the laser spot on the sample. The polarization of the reflected laser light was analyzed with a Wollaston prism and a standard optical bridge with balanced photodiodes. The polarization of the reflected light and the magnetic field were both measured and recorded simultaneously. To improve the signal-to-noise ratio, hysteresis loops were recorded and averaged over 50 field cycles. All MOKE measurements were performed at a temperature of 295 K.

**Magneto-transport device fabrication.** Electronic transport measurements were performed on multiple samples, with representative data presented in Extended Data Figure 4. Thin crystals were identified using an optical microscope, and electrode layouts were designed according to the shape of the target flake. Electrodes were defined via electron-beam lithography and lift-off processing, with contacting material being a 1 nm Ti adhesion layer and 40 nm Au contact metallization. Similar results were obtained with a metallization of 30 nm of Pt.

# Room Temperature Magnetic Order in Air-Stable Ultra-Thin Iron Oxide

Jiangtan Yuan[1#], Andrew Balk[2#], Hua Guo[1], Sahil Patel[1], Xuanhan Zhao[3], Qiyi Fang[1], Douglas Natelson[3], Scott Crooker[2], Jun Lou[1*]

[1]Department of Materials Science and NanoEngineering, Rice University, Houston, TX 77005, USA
[2]National High Magnetic Field Laboratory, Los Alamos, New Mexico 87545, USA
[3]Department of Physics and Astronomy, Rice University, Houston, TX 77005, USA
[#] Contributed equally to this work

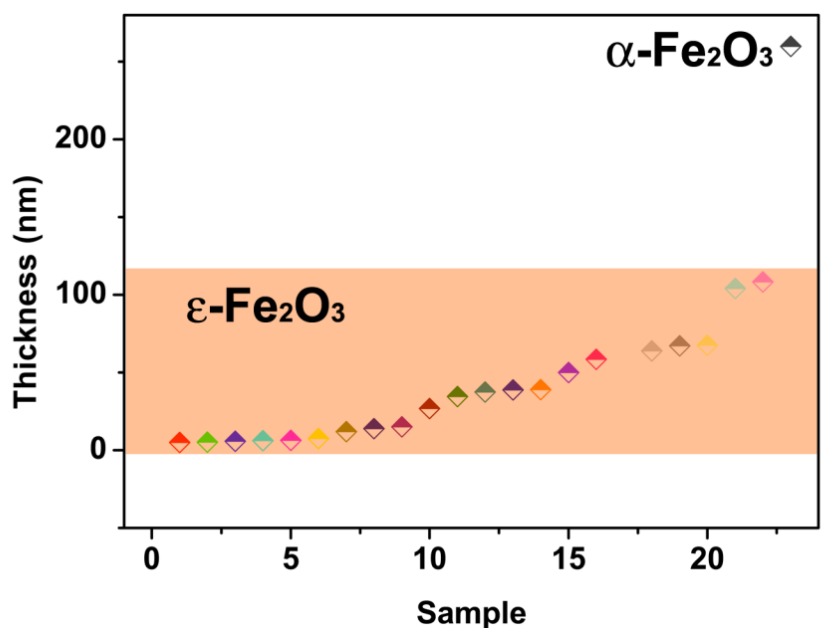

**Extended Data Figure 1 | Phase identification by Raman spectroscopy.** 22 out of 23 thin crystal samples measured on the same substrate are ε phase. Only the thickest is α–$Fe_2O_3$, suggesting that $Fe_2O_3$ preferentially forms into a stable ε phase in 2D, unlike in bulk.

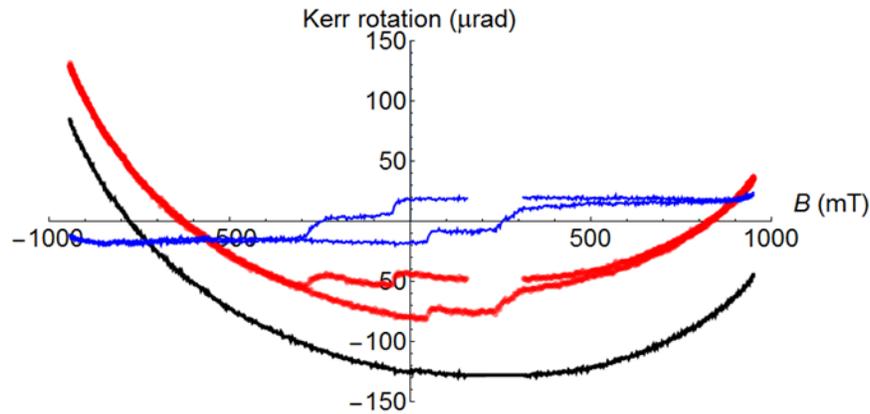

**Extended Data Figure 2 | Background removal from Kerr rotation measurements.** Motion of the sample during measurement and Faraday effect in the optics lead to non-hysteretic backgrounds in the raw Kerr data (red). We remove these backgrounds for more accurate data analysis. We first generate a model of the background by selecting subsets of the loop with decreasing absolute value of field $B$. We then adjusted the offset of these subsets to match each other through $B = 0$ (black). Then we fit this background to an empirical fifth order polynomial to model the background, and subtract this polynomial from the raw data to generate a background-free loop (blue). All subsequent analysis is performed on these background-free loops.

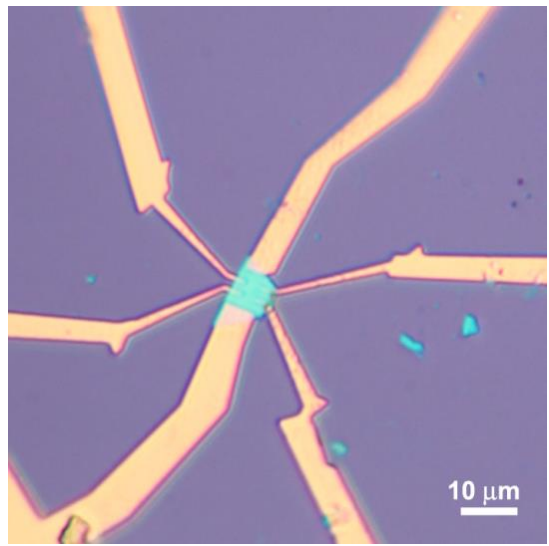

**Extended Data Figure 3 | Device geometry for magnetotransport measurement.** a $Fe_2O_3$ crystal approximately 10 nm thick with electrodes configured for a Hall measurement.

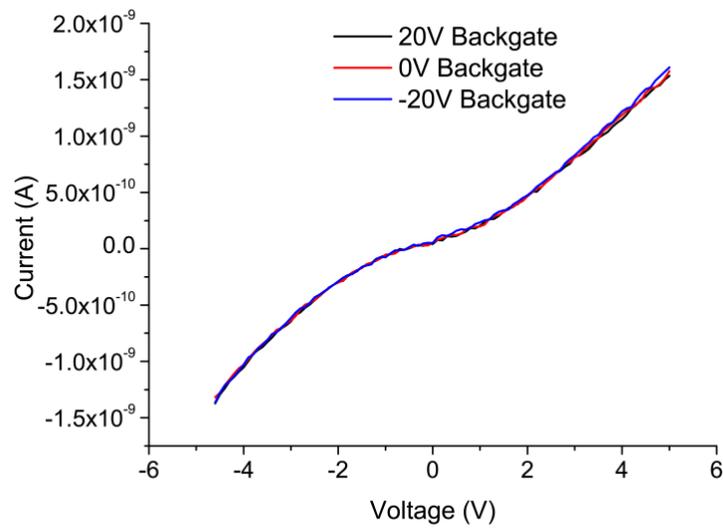

**Extended Data Figure 4 | Electronic transport measurement**. Two- and four-terminal measurements at room temperature revealed the samples to be highly resistive (resistivity ~ 100 Ω-m at room temperature), with no discernible Hall signal. Using the doped Si substrate (300 nm thermal oxide coating) as a backgate, further measurements showed no significant gate response. Neither a clear magnetoresistance nor a measurable Hall response were discernable.